\documentclass[aps,prb,reprint,groupedaddress]{revtex4-2}

\usepackage{graphicx}

\begin{document}



\title{How High a Field Can Be and Has Been Achieved in Superconducting Bulk Niobium Cavities Across Different RRR Values?}


\author{Takayuki Kubo}
\email[]{kubotaka@post.kek.jp}
\affiliation{High Energy Accelerator Research Organization (KEK), Tsukuba, Ibaraki 305-0801, Japan}
\affiliation{The Graduate University for Advanced Studies (Sokendai), Hayama, Kanagawa 240-0193, Japan}



\begin{abstract}
This Brief Note explores the relationship between residual resistivity ratio (RRR) and the maximum surface magnetic field in superconducting bulk niobium (Nb) cavities. Data from the 1980s to 2020s, covering RRR values from 30 to 500, are compared with theoretical performance limits, including the lower critical field ($B_{c1}$), superheating field ($B_{\rm sh}$), and thermal runaway field ($B_{\rm run}$). The results show that modern Nb cavities are approaching $B_{\rm run}$ and the metastability region above $B_{c1}$ across the entire RRR range but remain below the fundamental limit at $B_{\rm sh}$. Achieving $B_{\rm sh}$ requires not only advanced high-gradient surface processing but also improved thermal stability with low surface resistance, ultra-pure Nb, and optimized Kapitza conductance to ensure $B_{\rm run} > B_{\rm sh}$.
\end{abstract}

\maketitle



Superconducting radio-frequency (SRF) cavities accelerate particles using an electric field along the axis~\cite{Padamsee_textbook}. The accelerating gradient, $E_{\rm acc}$, is a key performance measure, as higher gradients shorten the accelerator length required for a given energy. However, the maximum $E_{\rm acc}$ is constrained by the material properties of the cavity.

The first limiting factor is the superconducting properties of the material, particularly the lower critical field ($B_{c1}$) and the superheating field ($B_{\rm sh}$)~\cite{Lin_Gurevich, 2020_Kubo, Kubo_erratum, Transtrum}. As $E_{\rm acc}$ increases, the peak surface magnetic field $B_0$ rises, where $B_0 = g E_{\rm acc}$, with $g$ set by the cavity design. Initially, the cavity remains in the Meissner state, but as the field increases, vortices penetrate, leading to RF losses and quenching. The Meissner state becomes metastable at $B_{c1}$, with an upper limit at $B_{\rm sh}$. Thus, the maximum achievable field, $B_0^{\rm (max)}$, is constrained within the {\it metastable band} between $B_{c1}$ and $B_{\rm sh}$. Both fields depend on the electron mean free path, tied to the residual resistivity ratio (RRR).

Another limitation comes from the material's thermal stability. Even without surface defects, like normal conducting residues, topographical irregularities, or weak superconducting precipitates, the exponential temperature dependence of the surface resistance $R_s$ creates a positive feedback loop~\cite{breakdown, Gurevich_Ciovati}. This feedback between the absorbed power, $(1/2)R_s H_0^2$, and the resulting temperature rise leads to {\it defect-independent thermal runaway} above a threshold field, $B_{\rm run}$. The threshold $B_{\rm run}$ depends on factors such as surface resistance, cavity wall thickness, thermal conductivity, and Kapitza conductance.

These fundamental limits, $B_{c1}$, $B_{\rm sh}$, and $B_{\rm run}$, can be enhanced by using high-purity niobium with a high RRR. Although the link between higher RRR and increased theoretical field is well known, a comprehensive summary of decades of cavity tests with varying RRR values remains unavailable.

This Brief Note compiles data~\cite{1984_Padamsee, 1987_Padamsee, 1995_Kako, 1997_Kako, 1999_Shishido, 2005_Kneisel, 2007_Brinkmann, 2007_Geng, Padamsee_textbook_2009, 2011_Singer, 2014_Kubo_LG, 2016_Ciovati, 2017_Shimizu, 2023_Howard} from the 1980s to 2020s, with RRR values ranging from 30 to 500, to investigate the relationship between RRR and the maximum achievable fields in Nb cavities. The results are compared with rough theoretical estimates of the fields as a function of RRR. This work establishes the foundation for discussing field limits in bulk Nb technologies, including both simple bulk Nb and multilayer approaches with thin films deposited on Nb~\cite{2006_Gurevich, 2014_Kubo, 2015_Gurevich, 2017_Kubo, 2021_Kubo, 2019_Antoine}.


\begin{figure*}[tb]
   \begin{center}
   \includegraphics[width=0.9\linewidth]{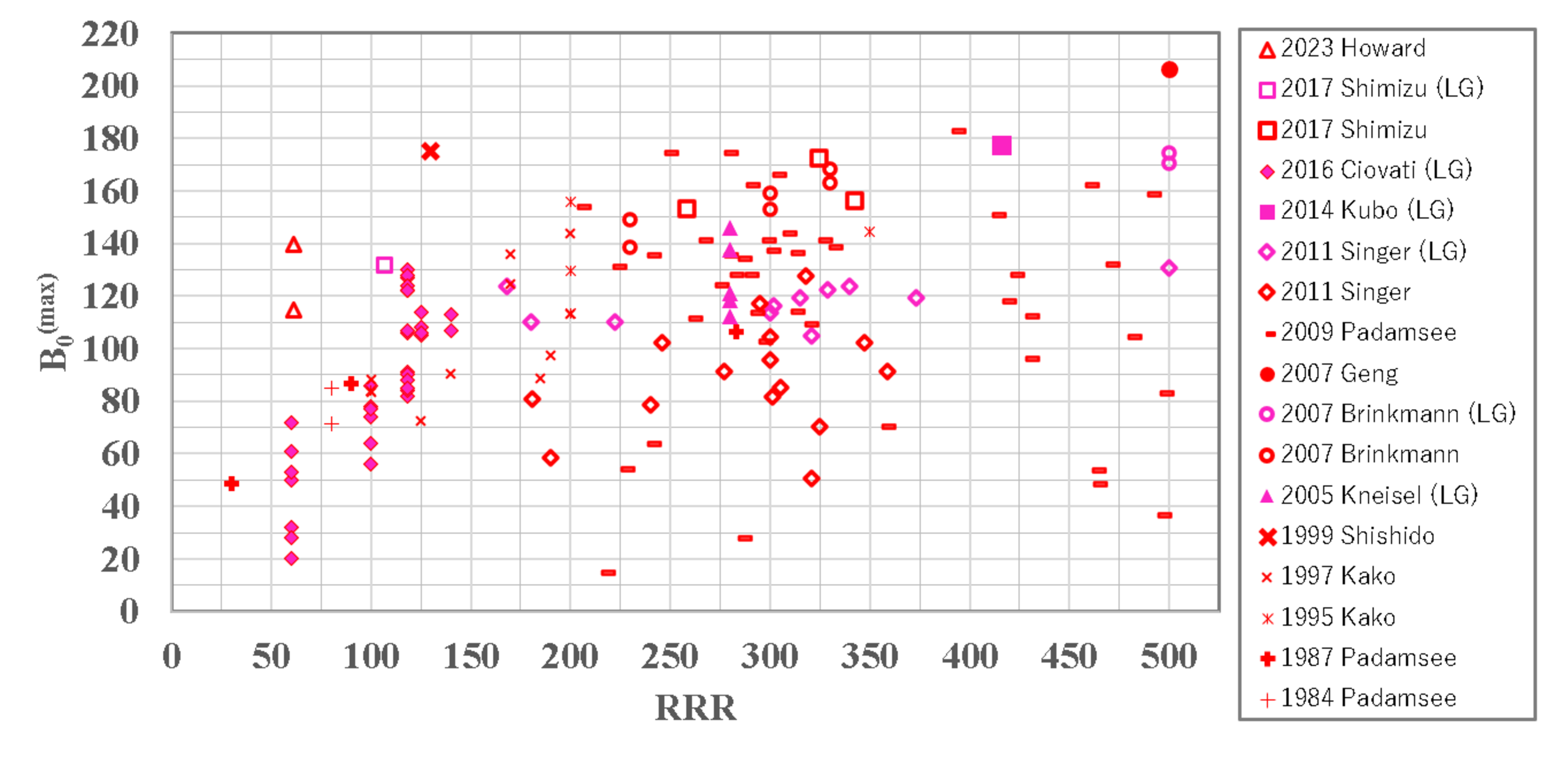}
   \end{center}\vspace{0 cm}
   \caption{
Maximum surface magnetic field $B_{0}^{\rm max}$ (mT) for $1.3$ and $1.5\,{\rm GHz}$ cavities as a function of RRR. The cavities were processed using chemical polishing and/or electropolishing, with or without low-temperature baking. The 147 data points are sourced from Refs.~\cite{1984_Padamsee, 1987_Padamsee, 1995_Kako, 1997_Kako, 1999_Shishido, 2005_Kneisel, 2007_Brinkmann, 2007_Geng, Padamsee_textbook_2009, 2011_Singer, 2014_Kubo_LG, 2016_Ciovati, 2017_Shimizu, 2023_Howard}, where RRR values were explicitly reported. Some high-performing cavities, such as those in Ref.~\cite{Grassellino}, which achieved $190\,{\rm mT}$, are not included due to the absence of RRR data.
   }\label{f1}
\end{figure*}

Figure~\ref{f1} compiles cavity test results of $B_0^{\rm (max)}$ for various RRR values collected over several decades~\cite{1984_Padamsee, 1987_Padamsee, 1995_Kako, 1997_Kako, 1999_Shishido, 2005_Kneisel, 2007_Brinkmann, 2007_Geng, Padamsee_textbook_2009, 2011_Singer, 2014_Kubo_LG, 2016_Ciovati, 2017_Shimizu, 2023_Howard}. The data show a broad range of $B_0^{\rm (max)}$, with some cavities achieving notable results ($\gtrsim 180\,{\rm mT}$). Cavities with ${\rm RRR} \simeq 200$ and even ${\rm RRR} \simeq 100$ reached $B_0 \simeq 150\,{\rm mT}$, corresponding to $E_{\rm acc} = 35\,{\rm MV/m}$ in Tesla-shaped cavities. However, concluding that ${\rm RRR} \sim 100$ is enough to consistently reach such gradients would be incorrect. The data do not reflect the success rate of cavities reaching these fields. For instance, the highest result for ${\rm RRR} \sim 130$, achieving $B_0^{\rm max} = 175\,{\rm mT}$, required 12 tests with repeated surface treatments~\cite{1999_Shishido}. It's important to note this figure shows the fields achieved in one or more tests without indicating the yield of high-performing cavities.

This dataset is compared with rough theoretical estimates of the maximum achievable fields ($B_{\rm sh}$, $B_{c1}$, $B_{\rm run}$) as functions of RRR, derived in the following.

$B_{\rm sh}$ represents the ultimate stability limit for the Meissner state, beyond which it becomes fully unstable. For dirty niobium, Ginzburg-Landau (GL) theory gives $B_{\rm sh}|_{T \simeq T_c} \simeq 0.745 B_c|_{T \simeq T_c}$ (see, e.g., Ref.~\cite{2017_Kubo} and references therein). As temperature decreases, this coefficient rises, with microscopic theory predicting $B_{\rm sh} \sim 0.8 B_c$ at $T\ll T_c$~\cite{Lin_Gurevich}. In the dirty limit, $B_{\rm sh} = 0.795 B_c \simeq 160\,{\rm mT}$ at $T=0$~\cite{2020_Kubo, Kubo_erratum, 2021_Kubo}. 
On the other hand, for clean niobium ($\kappa \sim 1$), GL theory predicts $B_{\rm sh}|_{T \simeq T_c} \simeq (1.2-1.3) B_c|_{T \simeq T_c}$~\cite{Pade2}, and extrapolating to $T\ll T_c$ yields $B_{\rm sh} \simeq (1.2-1.3) B_c \simeq 240-260\,{\rm mT}$. Since $B_{\rm sh}$ at low temperatures remains uncertain in most cases except for dirty niobium~\cite{Lin_Gurevich, 2020_Kubo, 2021_Kubo}, we adopt the standard practice of extrapolating the GL-based formula. We use Christiansen's formula~\cite{Christiansen}:
\begin{eqnarray} 
B_{\rm sh} (\kappa) = \frac{\sqrt{5}B_c}{3} \biggl( 1 + \frac{0.731}{\sqrt{\kappa}} \biggr), \label{Bsh}
\end{eqnarray}
originally derived for $\kappa > \kappa_c$ (where $\kappa_c \sim 1.1$ separates one- and two-dimensional critical perturbations). Numerical calculations~\cite{Transtrum} show this formula remains accurate for $B_{\rm sh}$ even at $\kappa \gtrsim 0.6$. The Pade approximation~\cite{Pade2}, derived for $\kappa < \kappa_c$, gives similar results to Eq.~(\ref{Bsh}) in the range $0.6 \lesssim \kappa \lesssim 1$.

$B_{c1}$ represents the field at which the Meissner state becomes metastable. To estimate $B_{c1}$, we use the approximate GL expression $B_{c1} = f(\kappa) B_c$~\cite{2003_Brandt}, where $f(\kappa) = \{\ln \kappa + C(\kappa)\}/\sqrt{2}\kappa$, and $C(\kappa) = 0.5 + (1 + \ln 2)/(2\kappa - \sqrt{2} + 2)$. Here, $B_c$ denotes the thermodynamic critical field. This formula is chosen for its balance between computational simplicity and accuracy. 
Though derived from GL theory and technically valid near $T_c$, the formula can be rewritten to provide a good approximation even at low temperatures.
Since $B_c$ is independent of nonmagnetic impurity concentration (Anderson's theorem), we get $B_{c1}(\kappa)/B_{c1}^{\rm clean}=f(\kappa)/f(\kappa_{\rm clean})$ or 
\begin{eqnarray}
B_{c1}(\kappa) = \frac{f(\kappa)}{f(\kappa_{\rm clean})} B_{c1}^{\rm clean}. \label{Bc1}
\end{eqnarray}
Here, $B_{c1}^{\rm clean}$ and $\kappa_{\rm clean}$ refer to pure niobium. Using $B_{c1}^{\rm clean} = 185\,{\rm mT}$ at $T=0$ and $\kappa_{\rm clean} = 0.8$ for Nb~\cite{1969_Ikushima}, this approximation matches experimental data for $B_{c1}$ at $T=0$ across $\kappa < 1.5$~\cite{1969_Ikushima}, fitting our range of interest (as discussed, $\kappa < 1.5$ corresponds to ${\rm RRR} \gtrsim 20$). Hence, extrapolating this formula to lower temperatures appears effective.

The GL parameter $\kappa$ is related to the mean free path $\ell$ via microscopic theory~\cite{1959_Gorkov}:
\begin{eqnarray}
\kappa (\ell) = \frac{1}{\chi(a_{\rm imp})} \kappa_{\rm clean}, \label{kappa} 
\end{eqnarray}
where $\chi$ is the Gor'kov function, 
\begin{eqnarray}
&&\chi (a_{\rm imp}) = \frac{8}{7\zeta(3)} \sum_{n=0}^{\infty} \frac{1}{(2n+1)^2 (2n+1+a_{\rm imp}) }, \\ 
&&a_{\rm imp} = \frac{\pi }{2e^{\gamma_E} } \frac{\xi_0}{\ell} \simeq 0.882 \frac{\xi_0}{\ell}, \\ 
&&\kappa_{\rm clean} = \frac{2e^{\gamma_E}}{\pi}\sqrt{\frac{6}{7\zeta(3)}} \frac{\lambda_0}{\xi_0} \simeq 0.957 \frac{\lambda_0}{\xi_0}.
\end{eqnarray}
Here, $\zeta$ is the Riemann zeta function, $\gamma_E = 0.577$ is Euler's constant, $\xi_0=\hbar v_f/\pi\Delta_0$ is the BCS coherence length, and $\lambda_0^{-2}=(2/3)\mu_0 N_0 e^2 v_f^2$ is the BCS penetration depth in the clean limit at $T=0$, with $v_f$ as the Fermi velocity and $N_0$ the normal electron density of states at the Fermi energy. The parameter $a_{\rm imp}$ characterizes the material's dirtiness.
The mean free path $\ell$ is expressed in terms of RRR as $\ell = (3.7 \times 10^{-16}\,{\rm \Omega \cdot m^2})/\rho_n$, where $\rho_n \simeq \rho_n(295\,{\rm K})/ {\rm RRR}$ and $\rho_n(295\,{\rm K})= 1.45 \times 10^{-7} \,{\rm \Omega \cdot m}$. This leads to:
\begin{eqnarray}
\ell = {\rm RRR} \times 2.55\,{\rm nm}. \label{ell}
\end{eqnarray}
These formulas [Eqs.~(\ref{Bsh})-(\ref{ell})], along with empirical temperature dependencies of $(B_{c1}^{\rm clean}, B_c ) \propto 1-(T/T_c)^2$ with $T_c=9.2\,{\rm K}$, form the poor man's formula for calculating $B_{sh}$ and $B_{c1}$ as functions of RRR (or $\ell$) at any temperature.

\begin{figure}[tb]
   \begin{center}
   \includegraphics[width=0.9\linewidth]{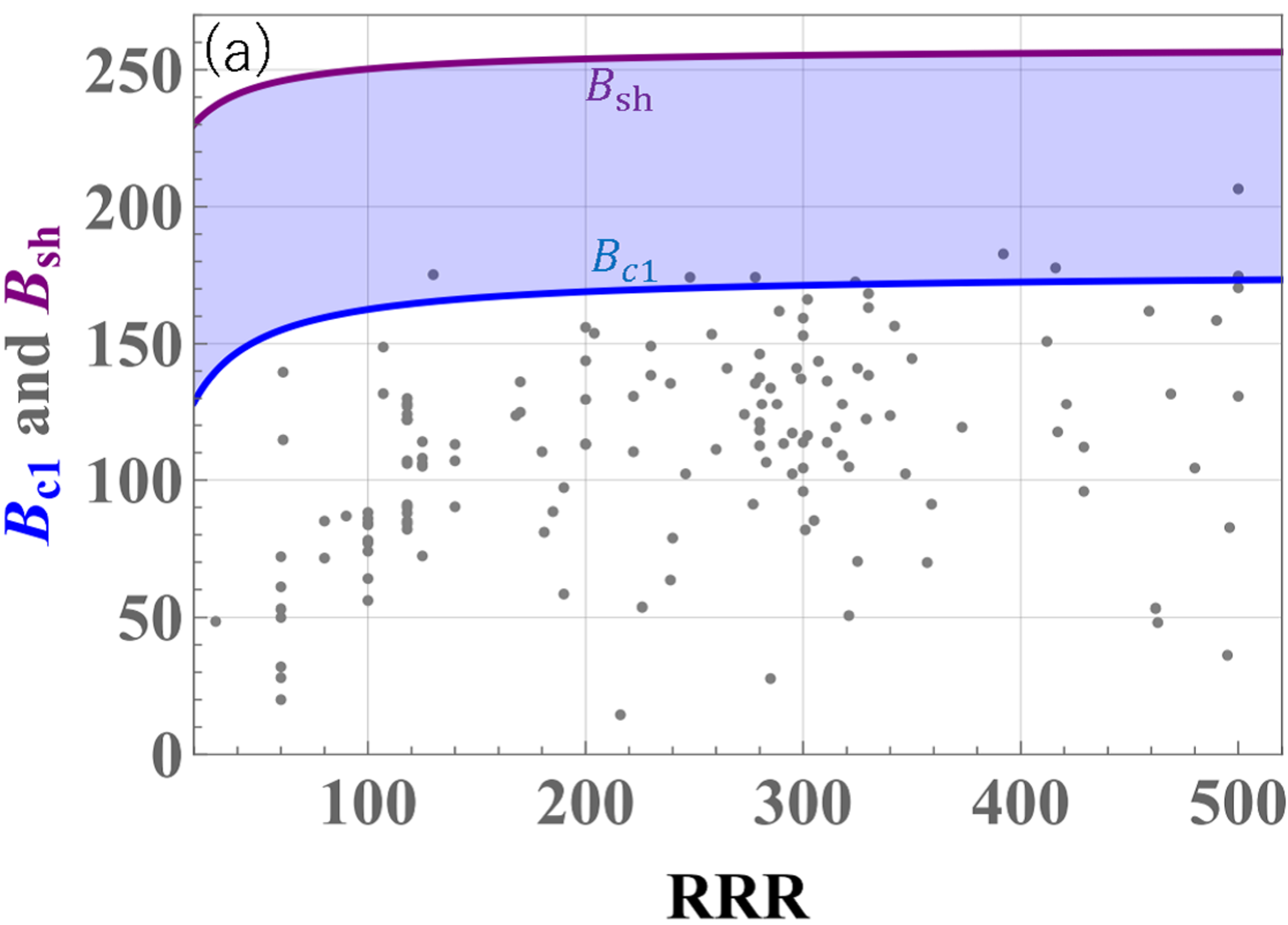}
   \includegraphics[width=0.9\linewidth]{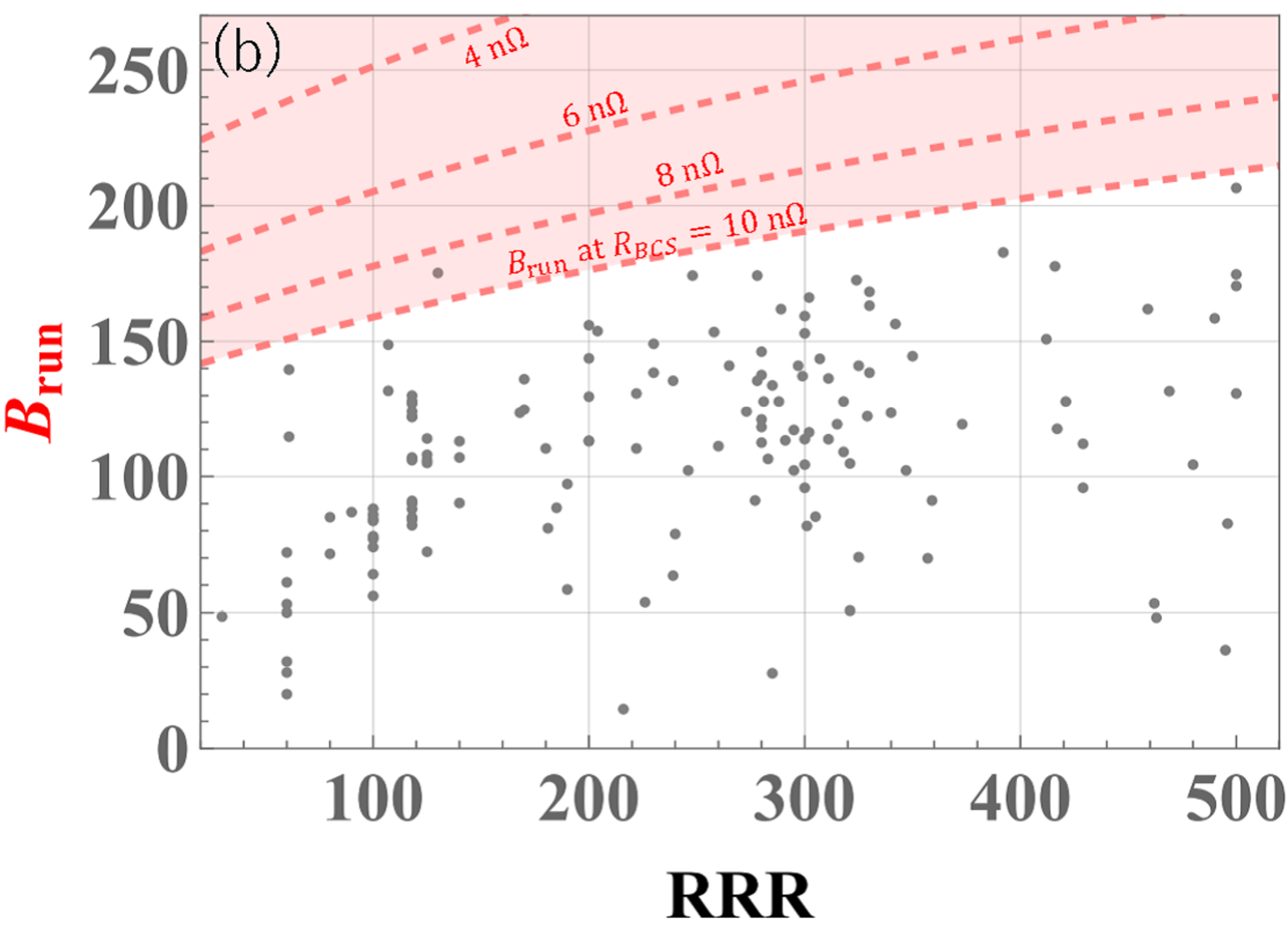}
   \end{center}\vspace{0 cm}
   \caption{
(a) $B_{\rm sh}$ (purple) and $B_{c1}$ (blue) at $2\,{\rm K}$ as functions of RRR, estimated using Eqs.~(\ref{Bsh})-(\ref{ell}).
(b) $B_{\rm run}$ (red) at $2\,{\rm K}$ as a function of RRR, calculated for $R_{\rm BCS}(2\,{\rm K}) \leq 10\,{\rm n\Omega}$. 
The 147 data points from Fig.~\ref{f1} are overlaid (gray points) in both figures for comparison.
   }\label{f2}
\end{figure}

Fig.~\ref{f2} (a) shows $B_{\rm sh}$ (purple) and $B_{c1}$ (blue) as functions of RRR at $2\,{\rm K}$. The input parameters are $B_{c1}^{\rm clean}|_{T=0} = 185\,{\rm mT}$, $\kappa_{\rm clean} = 0.8$, and $\xi_0 = 45\,{\rm nm}$~\cite{1969_Ikushima}. The blue area between the curves represents the metastability region, beyond which the Meissner state becomes completely unstable. Although many cavities surpass $B_{c1}$ and enter this region, none have yet reached $B_{\rm sh}$.

$B_{\rm run}$, the threshold field for {\it defect-independent thermal runaway}, is another critical limit for superconducting cavities. This breakdown field~\cite{breakdown, Gurevich_Ciovati} is analyzed using the heat balance equation for the inner surface temperature. To focus on the theoretical field limit, we assume the residual surface resistance, which lowers $B_{\rm run}$, is negligible compared to the thermally activated quasiparticle contribution (i.e., \( R_{\rm res} \ll R_{\rm BCS} \)), a condition approximately met in well-controlled $2\,{\rm K}$ experiments, especially when the magnetic environment is optimized to minimize trapped flux, a key source of residual resistance (see, e.g., Ref.~\cite{Geng_flux} and references therein). Under these conditions, $B_{\rm run}$ is expressed as~\cite{breakdown, Gurevich_Ciovati}:
\begin{eqnarray}
B_{\rm run} = 
\mu_0 \sqrt{ \frac{2 T/T_c }{\alpha e } \frac{T}{r R_{\rm BCS}} }, 
\label{Brun}
\end{eqnarray}
where \( e = 2.718 \) is Napier's constant, \( \alpha = \Delta/k T_c  \simeq 1.9 \) for Nb, \( r(T, {\rm RRR}) = h_{\rm K}^{-1} + d/K \) is the thermal resistance, \( K \) is the thermal conductivity, and \( h_{\rm K} \) is the Kapitza conductance. The BCS surface resistance is given by \( R_{\rm BCS} = (A/T) e^{-\Delta/k T} \), where $A$ is a constant. Although \( R_{\rm BCS} \) depends on the field~\cite{2014_Gurevich, Kubo_Gurevich}, we do not account for this in the current analysis. Instead, we vary \( A \) to estimate $B_{\rm run}$ with associated uncertainties.

The thermal conductivity \( K \) depends on RRR. For our calculations, we set \( d = 2.8\,{\rm mm} \) and use the formula from Ref.~\cite{Bonin} to determine \( K({\rm RRR}) \), applying the room temperature resistivity $\rho_n (295\,{\rm K}) = 1.45\times 10^{-7}\,{\rm \Omega \cdot m}$ and the phonon mean free path $\ell_{\rm ph} = 75\,{\rm \mu m}$. For the Kapitza conductance \( h_{\rm K} \), we adopt \( h_{\rm K} = 5000\,{\rm W\, m^{-2}\, K^{-1}} \) from Ref.~\cite{Kapitza}. With these values, the thermal resistance (\( r = h_{\rm K}^{-1} + d/K \)) equals approximately \( 7\,{\rm K \, m^2 \, W^{-1}} \) at $2\,{\rm K}$ for \( {\rm RRR} = 300 \). This estimate aligns well with recent measurements of thermal resistance in the range of \( 4\text{-}6\,{\rm K \, m^2 \, W^{-1}} \) at $2\,{\rm K}$ for \( {\rm RRR} > 250 \)~\cite{Dhakal}.

Fig.~\ref{f2} (b) shows $B_{\rm run}$ at $2\,{\rm K}$ as a function of RRR. Although the absolute value of $B_{\rm run}$ contains some uncertainty due to variations in the thermal resistance $r$, and $R_{\rm BCS}$ generally exhibits intrinsic field dependence, ranging from $1\,{\rm n\Omega}$ to $10\,{\rm n\Omega}$~\cite{2014_Gurevich, Kubo_Gurevich}, which introduces some imprecision in $B_{\rm run}$ within the red region of the figure, the overall trend suggests that $B_{\rm run}$ remains on the same order of magnitude as $B_{c1}$ and $B_{\rm sh}$.

To achieve $B_{\rm sh}$, it is essential to ensure that $B_{\rm run}$ exceeds $B_{\rm sh}$. If $B_{\rm run}$ remains below $B_{\rm sh}$, thermal runaway will occur before reaching the superheating field, posing an insurmountable barrier. Achieving this will require not only advanced high-gradient surface processing but also enhanced thermal stability, characterized by low surface resistance, adequate thermal conductivity (i.e., bulk RRR), and optimized Kapitza conductance to guarantee $B_{\rm run} > B_{\rm sh}$.

Before concluding, it's important to note the role of defects. Cavity performance is impacted by resistance to local heating at defect sites, governed by thermal conductivity $K$, which depends on RRR. Defects acting as heat sources cause temperature spikes, but higher RRR materials mitigate these, enabling higher fields. This drove efforts in the 1980s and 1990s to raise niobium's RRR from reactor-grade (${\rm RRR} \simeq 20$) to over 300. This improvement in RRR and its impact on cavity performance is well documented during this period.


\begin{acknowledgments}
I am deeply grateful to those who supported my three-year paternity leave~\cite{ikuji}, during which I revisited an unfinished project initiated in 2018 for the TESLA collaboration meeting held at RIKEN, leading to the development of this Brief Note. This work was supported by JSPS KAKENHI Grants No. JP17KK0100 and Toray Science Foundation Grants No. 19-6004.
\end{acknowledgments}

\end{document}